# Multi Agent Communication System for Online Auction with Decision Support System by JADE and TRACE

A. Martin[*1], T. Miranda Lakshmi[*2], J. Madhusudanan[*3]

[*1] Dept. of Master of Computer Applications,
Sri Manakula Vinayagar Engg. College, Puducherry. jayamartin@yahoo.com
[*2], Corresponding author Dept. of Computer Science & Applications,
St. Joseph's College, Cuddalore. cudmiranda@gmail.com
[*3], Corresponding author Dept. of CSE, Sri Manakula Vinayagar Engg. College,
Puducherry, India. contactmadhu@gmail.com
doi: 10.4156/jcit.vol4.issue2.martin

**Abstract**

*The success of online auctions has given buyers access to greater product diversity with potentially lower prices. It has provided sellers with access to large numbers of potential buyers and reduced transaction costs by enabling auctions to take place without regard to time or place. However it is difficult to spend more time period with system and closely monitor the auction until auction participant wins the bid or closing of the auction. Determining which items to bid on or what may be the recommended bid and when to bid it are difficult questions to answer for online auction participants. The multi agent auction advisor system JADE and TRACE, which is connected with decision support system, gives the recommended bid to buyers for online auctions. The auction advisor system relies on intelligent agents both for the retrieval of relevant auction data and for the processing of that data to enable meaningful recommendations, statistical reports and market prediction report to be made to auction participants.*

# 1. Introduction

An auction is an important type of business to sell products based upon effective pricing methods. It is a method of selling property in a public forum through open and competitive bidding by offering the item to the highest bidder. Online-Auctions are one of the successful types of electronic marketplaces. They bring together buyers and sellers on a massive scale. The online auction is one in which participants bid for products and services at any time and from anywhere in the world over the internet.This paper addresses four parts of multi-agent communication system.

- Intelligent Auction Agent
- Agent Based Decision Support System
- Statistical Report and Market Prediction Report
- Information Retrieval Agents

## 1.1. Intelligent Auction Agent

The first part is Intelligent Agents or Auto bid Agents – It is like a house broker, who seeks rented houses on behalf of us until to meet our requirements. Like a house broker in online auction one can appoint or configures software agents. This software agent places bids on behalf of buyers. The result of auction is informed to the auction participants.

## 1.2. Agent Based Decision Support System

The main objective of agent based decision support system is to provide recommended bid using agents. Using decision support system one can get appropriate bid value for his auction. Developing a decision support system is costly process; instead one can make use of the existing decision support system. In www.ebay.com a software agent is constructed and it is named as TRACE. This TRACE is connected with a decision support system. Agents can get recommend bid, by submitting item name, current bid, number of bidders, remaining auction duration to the TRACE. The TRACE made analysis with help of decision support system, and then finally submits the recommended bid. This recommended bid is intern submitted to the requested agent. In this paper an agent has been constructed using Java Agent Development Environment and it is called as JADE that is live in www.auctionagent.com. Thus JADE and TRACE communicates through Agent Communication Channel [ACC] and Agent Communication Language [ACL] and gives the recommended bid.

### 1.3. Statistical Report and Market Prediction Report

JADE Summarizes Minimum value, Median value and Maximum value for a product and Number of Quantities sold from the closed and current auctions. To find this statistics report an auction data warehouse has been constructed. By using regression analysis method the JADE made analysis and provides Market Prediction Report. In the recent past years for what price the product was sold and near future for what price the item may sell has been given as report from the auction data warehouse.

### 1.4. Information Retrieval Agents

To construct auction data warehouse auction data collection should be made. Data collection process can be done through JADE – whenever any item being sold in any website through auction that data can be collected by agents and stored in the auction database. XML parser may be used as data retrieval tool.

## 2. Related Work
### Agents & Intelligent Agents

- Agents - Systems that can decide for them selves what they need to in order to satisfy their design objective.
- Agent that must operate robustly in rapidly changing, unpredictable or open environments, where there is a significant possibility that actions can fail are known as intelligent agents or autonomous agents.
- E.g.:- A rocket moves from earth to outer planets. A ground crew is usually required to continually track its progress and decide how to deal with unexpected eventualities.

When do we consider an agent to be intelligent?

An intelligent agent is one that is capable of flexible autonomous action in order to meet its design objectives. The agent should possess the following flexible autonomous actions,

**Reactivity** - Intelligent agents are able to perceive their environment and responds in a timely fashion to changes that occur in it in order to satisfy their design objectives.

**Pro-activeness** - Able to exhibit goal directed behavior by taking the initiative in order to satisfy their design objectives.

**Social ability** - The agent is capable of interacting with other agents in order to satisfy their design objectives.

## 3. Statistical Support for Recommended Bid

For a buyer, placing of appropriate bid is very important. A Professor Dawn G. Gregg, Dept. of Information Systems, University of Colorado at DENVER has conducted a survey on auctions. On average the maximum price paid for an item was 79% higher than the average price paid and 36% higher than minimum price paid. This indicates that many buyers are unaware of the appropriate price for the products they are purchasing through online.

Other important factors,
- What is the appropriate time to place the bid?
- Numbers of bidders are currently participating in the auction.

Data was gathered for 10,522 auctions. In 38% of these auctions, the winning bid was placed in the last hour of bidding. In 26% of auctions the winning bid was a new bid, which is placed in the last 5 minutes of the auction. So, we need an excellent decision support system or expert advice or statistical report to find the appropriate bid value.

## 4. FIPA Agent Platform

The standard model of an agent platform, as defined by FIPA, is represented in the following figure1,

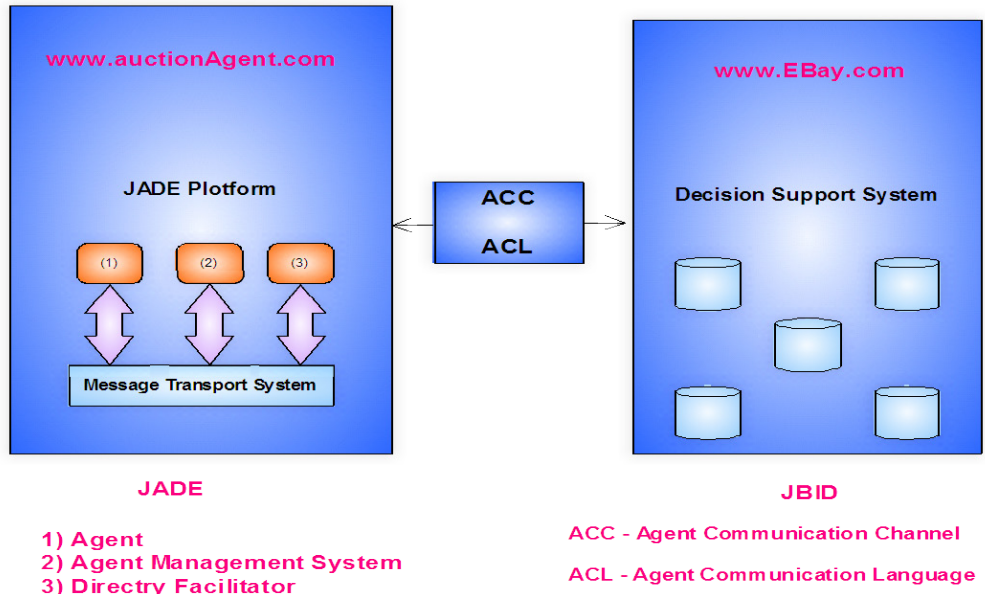

**Figure 1**, Reference Architecture of a FIPA Agent Platform

### 4.1. Agent Management System

The Agent Management System (AMS) is the agent who exerts supervisory control over access to and use of the Agent Platform. Only one AMS will exist in a single platform. The AMS provides white-page and life-cycle service, maintaining a directory of agent identifiers (AID) and agent state. Each agent must register with an AMS in order to get a valid AID. The Directory Facilitator (DF) is the agent who provides the default yellow page service in the platform.

### 4.2. Agent Communication Channel

The Message Transport System, also called Agent Communication Channel (ACC), is the software component controlling all the exchange of messages within the platform, including messages to/from remote platforms. JADE fully complies with this reference architecture and when a JADE platform is launched, the AMS and DF are immediately created and the ACC module is set to allow message communication. The agent platform can be split on several hosts. Only one Java application, and therefore only one Java Virtual Machine (JVM), is executed on each host. Each JVM is a basic container of agents that provides a complete run time environment for agent execution and allows several agents to concurrently execute on the same host.

The main-container, or front-end, is the agent container where the AMS and DF lives and where the RMI registry, that is used internally by JADE, is created. The other agent containers, instead, connect to the main container and provide a complete run-time environment for the execution of any set of JADE agents.

According to the FIPA specifications, DF and AMS agents communicate by using the FIPA-SL0 content language, the FIPA-agent-management ontology, and the FIPA-request interaction protocol. JADE provides compliant implementations for all these components. The following messages are used by agents to communicate with other agents,

a. Accept Proposal - One Agent accepts the proposal of another agent
b. Agree - The process of mutual agreement between agents for request and response
c. Inform - One agent informs to the agent about the process
d. Failure - The communication failure is send as the response
e. Propose - One agent submits some task to the other agent
f. Refuse - The process of denying some task
g. Request - Request for some action to be performed

## 5. Multi-Agent Communication Architecture

The Auction Advisor System follows the Multi-tier Architecture. The Agent TRACE also called as JBID or TRACKER. The below figure explains the architectural design for multi-agent communication. The details of various tiers present in this system is given below,

### 5.1. The Data Layer

It consists of the auction data that are necessary for the functioning of the auction advisor system. There are three primary sources of data for the system,

- The Web contains the data on current and recently closed auctions.
- The Auction database contains transactional auction data.
- A data warehouse contains the historic auction data previously collected by Information Retrieval agents

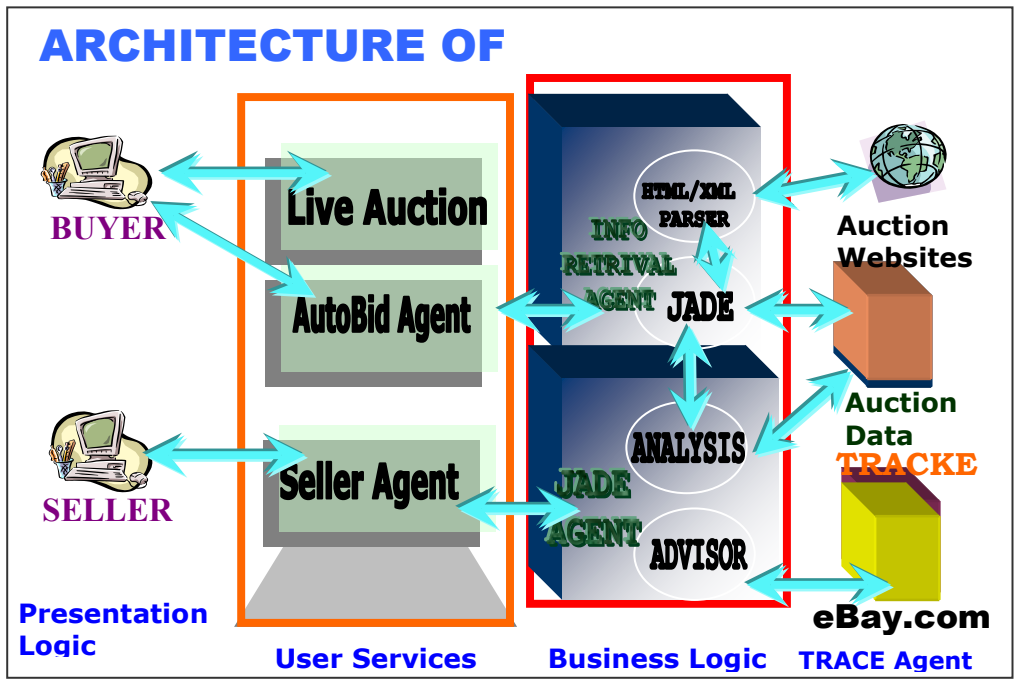

**Figure 2**, Architectural design for Multi-Agent Communication for Agent Based Decision Support System.

### 5.2. Business Logic Layer

- It contains the agents responsible for retrieving, persist the auction data and processing/analyzing the online auction data.
- These agents communicate with buyer and seller process in the User Service layer.
- An information retrieval agent responsible for parsing of auction data [XML files] from various auction servers and stores the data in the auction database.
- JADE made analysis and prepares Auction Statistical Report and Market Prediction Analysis.
- JADE made communication with eBay.com and it's associated Decision Support System to collect Expert Advice on Recommended Bid.

### 5.3. User Services Layer

- This layer provides services to buyers and sellers.
- Buyer can place bids manually by using live auction.
- They appointment agents to place bid on behalf them.

### 5.4. Presentation Layer

- The Presentation layer consists of the user interfaces that enable auction agents to communicate and interact with users.
- User to agent communication is required to enable the agent to perform auction bidding.

## 6. Experimental Design and Implementation Software Architecture

The following figure 5.2 explains the software architecture for auction advisor system, in the client side the presentation layer is developed using ASP.Net. In serve side the live auction is developed with VB.net and AutoBid agents developed using C#.net. Agent has been constructed using Java Agent Development Environment [JADE] and it is able to communicate with another agent called as TRACE lives in eBay.com. Microsoft XML Parser is used to parse the XML auction files. To avoid performance degradation auction data is stored as XML files. MS-Access is used as Database.

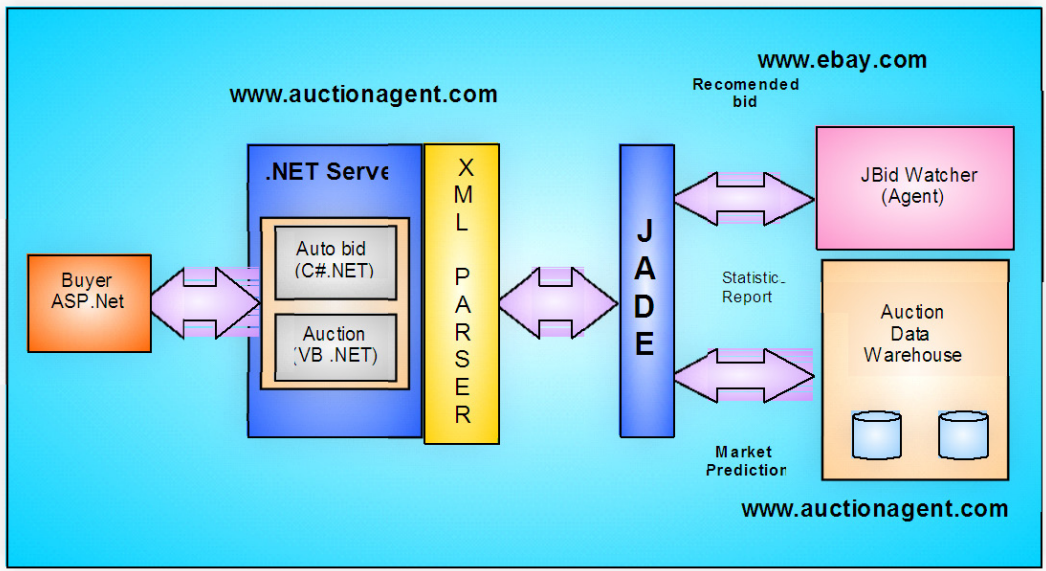

**Figure 3**, Software Architectural design for Intelligent Auction

### 6.1. JADE and TRACE Agents Communication

Once the buyer is authenticated, he has been taken into live auction. The buyer may select any product. Expert advice i.e., recommended bid or appropriate bid is provided to all the products. Once the expert advice has been selected a XML file is created with the following details.
- Item Name
- Current Bid
- Number of Bids and
- Remaining auction duration.

The specific requirement of XML file is that, in near future JADE may be moved to some other server. It gives flexibility to transfer the data from one server to another server. The XML file is parsed using Microsoft Parser and data is transferred to JADE. Intern it is forwarded to TRACE which is live in eBay.com. TRACE is connected with Decision Support System. TRACE made analysis and returns the intelligent bid or recommended bid. The above steps are carried out in backward direction and buyer gets recommended bid.

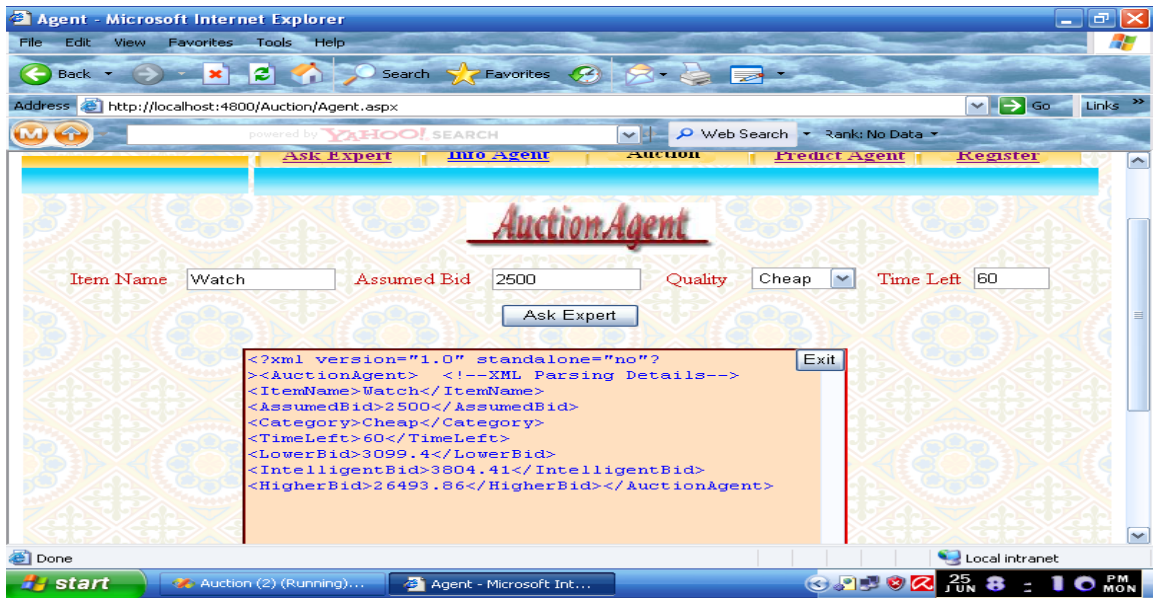

**Figure 4,** Auction Agent screen, this screen is used for asking the expert advice for a particular product.

Placing of bid is branched into two parallel activities. One is LiveBid and other is AutoBid. In LiveBid the bid is placed manually by action participants and in LiveBid the bid is placed by agents according to buyer configuration. Among all the bids the maximum bid would be found out and it is declared as winning bid at that moment. Once auction duration reaches one minute before the auction to end, it checks for a new auction. If any new bid is available then auction duration extends for another one minute. The above steps repeated until the placing of any new bid, otherwise the Maximum bid is find out and it is declared as winning bid.

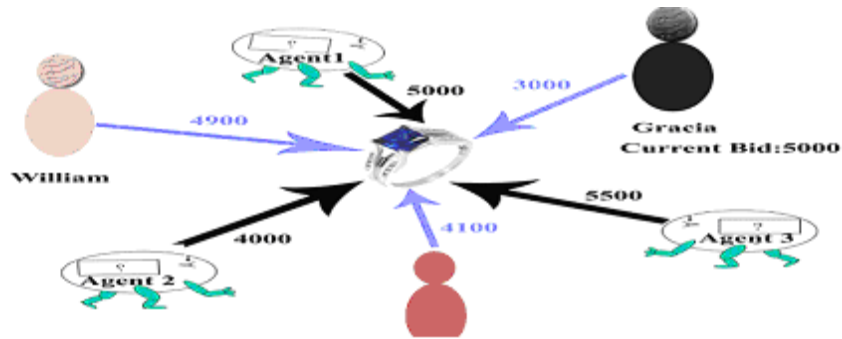

**Figure 5,** declaring the winning bid

### 6.2. How one AutoBid compete with other LiveBids and AutoBids

Once the bidder creates autobid it checks for the current bid. If the current bid outbids the autobid, the autobid places a new bid. It tries to maintain its bid as winning bid until it reaches buyer specified maximum value with bid increment. At a particular moment the AutoBid is having three choices,

When a placed bid becomes outbid, there are two options,
  a. Set a new Maximum Bid
  b. Cancel the AutoBid

When it becomes the Maximum Bid, the AutoBid Compete with other bids and tries to be a winning bid.

### 6.3. Market Prediction Report

By using regression analysis the JADE gives the Market Prediction Report. In the recent past years for what price the product was sold and near future for what price the item may sell has been given as report from the auction data warehouse.

Regression formula,
1. Variant1 = int (past * random()) + present.
2. Variant2 = (past + present)/2.
3. Future = int (Variant2 + random()) + Varinat1.

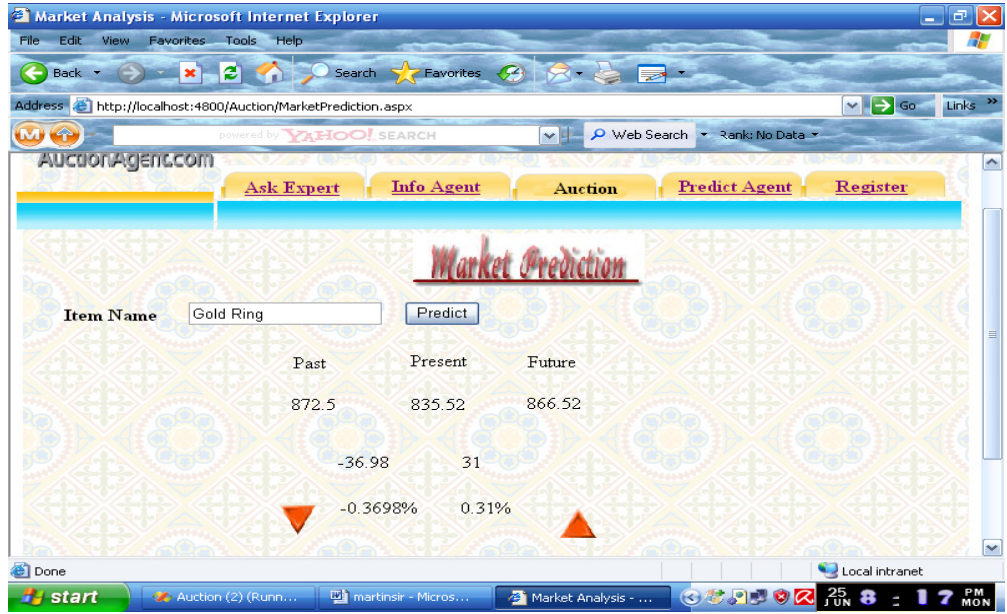

**Figure 6,** JADE made analysis and gives Market Prediction Report.

### 6.4. Information Retrieval Agents

To construct auction data warehouse auction data collection should be made. Data collection process may be done through JADE – whenever any item being sold in any website through auction that data can be collected by agents and stored in the auction database. XML parser may be used as data retrieval tool. In future agents trained to learn the domain structure of any web site, such that any publicly available data can be extracted by the agents. From auction operational database a data warehouse has been constructed. By applying formula market prediction analysis report is prepared.

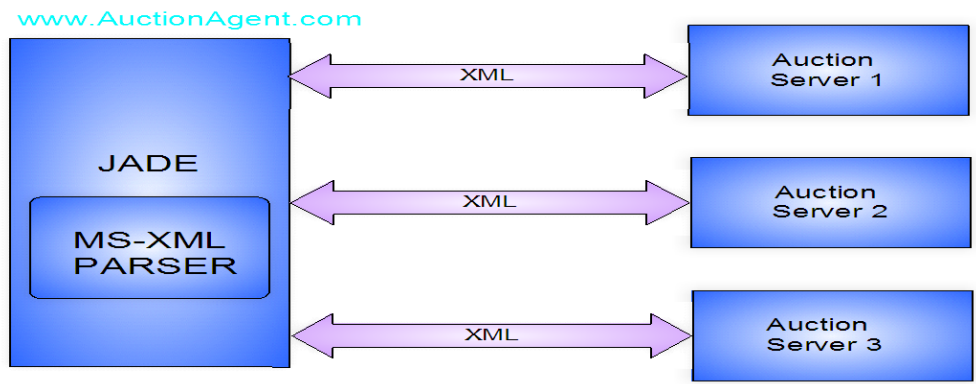

Figure 7, Information Retrieval by Software Agents

## 7. Experimental Results and Agents Performance

The recommended bid given by TRACKER and JADE is summarized in the following tables,

**Table 1** Recommended bid given by TRACE which is live in www.ebay.com

| Auction Site | Title | Current Bid | Recommended Bid | Winning Bid |
|---|---|---|---|---|
| EBay | Sony Digital Camera C-3020 | Rs.7252 | Rs.7353 | Rs.7330 |
| BidZ | Sony Digital 2.0 Mega Pixel Camera | Rs.8312 | Rs.8353 | Rs.8397 |
| EBay | Sony Digital Kit Camera C-3027 | Rs.6298 | Rs.6353 | Rs.6323 |
| EBay | Sony Digital Original Packing Camera C-3028 | Rs.8190 | Rs.8353 | Rs.8333 |
| EBay | Sony Digital Camera C-3021 | Rs.7280 | Rs.7353 | Rs.7315 |
| BidZ | Sony Digital Camera C-3020 | Rs.6154 | Rs.6353 | Rs.6350 |
| MyFind | Sony Digital Zoom Camera C-3020 | Rs.9300 | Rs.9353. | Rs.9360 |

**Table 2** Recommended bid given by JADE which is live in www.auctionagent.com

| Auction Site | Item Code | Title | Current Bid | Recommended Bid | Winning Bid |
|---|---|---|---|---|---|
| auction agent .com | 10747 | Ring With 2.30ctw Precious Stones 8.2g - Size 7.5. | Rs.17500 | Rs.18700 | Rs.19000 |
| | 13378 | Dazzling Ring Made of Solid 14K White Gold - Size 7.25. | Rs.18890 | Rs.19200 | Rs.19000 |
| | 17387 | Gents Ring With weight 10.0g - Size 10. | Rs.16400 | Rs.16700 | Rs.18000 |

The above said two tables summarizes the recommended bid various auction items. In most of the auctions the recommended bid given by agents is 75% matches with the winning bid. In very few auctions only the deviation between recommended bid and Winning bid is high.

## 8. Future Work

In future the system is developed with the following features,

### 8.1. Predicting the changing auction market behavior using data mining agent

The Auction Advisor agent system can be used by data-mining agent to support online auction market development. The following steps are required to execute it,

1. The first step is the information retrieval agent should catch publicly available auction data. To catch the publicly available data the IRA should learn the domain knowledge of each publicly available auction data. The collected auction data should be parsed and verified whether it is auction data using information retrieval agents.
2. The second step is construction auction data warehouse. The present and previously collected data should be added to data warehouse.
3. The third step is construction of Rule Base
   - It should contain all the auction rules.
   - New auction rule should be developed using action data warehouse.
   - Some of the auction rules are…,

**The buyer heuristics are (from the literature),**

> Bids need not exceed the recent median sales price on recently closed auctions.
>
> *Recommended_maximum_bid = median_closed_price*
>
> Bidders should place bids towards the end of the auction duration.
>
> *Recommended_bid_time = close_time – 5 minutes*
>
> Bidders should track/participate in auction with fewer bids/bidders.
>
> *if (numbers<Median_num_bids) bid = true;*
>
> *else bid = false;*

- Finally Data warehouse, Rule base and Data mining agent should be combined to find the changing or new auction customer behaviors.

### 8.2. Short Message Service

Short Message Service (SMS) message will be send to the auction participants. It will definitely help the auction participants to be informed about the status of the auction and the arrival of the new products.

### 9. Conclusion

The Agent Based Decision Support System can be used to provide price histories and other auction information help to educate auction participants. In the dynamic world of online auctions the real value of an item is constantly changing, hence there is a significant need for current data, so that the appropriate current price (or valuation) can be determined.

One of the benefits provided by the multi-agent communication auction advisor system is that it continuously retrieves the most recent data available when performing data analysis and making recommendations to auction participants. Buyers using the auction advisor know what a reasonable price for a given item in which can be avoid paying too much.

Users can use the autobid agents to select several items to bid on and to time their bids to maximize their probability of winning a given item within a specified period of time. Buyers could use the statistical information to determine the appropriate least, average and highest prices for their auctions.

Introducing agents into online auctions have fundamentally changed the way the auctions operate and the outcomes for both buyers and sellers. As online-auction buyers and sellers increase their use of intelligent agents to automate information gathering and consequently auction decisions. Utilization of this auction advisor like agents can keep the electronic consumers better informed of current trend of market environment and maximum outcome for consumers.

### 10. References

[1]. Ira Rudowsky, "Intelligent Agents", *Proceedings of the Americas Conference on Information Systems*, New York, August 2004.

[2]. Morozov and Yu. V. Obukhov, An Approach to Logic Programming of Intelligent Agents for Searching and Recognizing Information on the Internet, Pattern Recognition and Image Analysis, Vol. 11, No. 3, 2005, pp. 570–582.

[3]. D. Gregg and S. Walczak, " Auction advisor: an agent-based online-auction decision support system ", *The International Journal of Decision Support Systems,* Volume 41, Issue 2, Pages: 449 – 471, 2006.

[4]. Maria Fasli and Michael Michalakopoulos, Developing Software Agents using .NET, Department of Computer Science, University of Essex, UK.

[5]. Greenwald, and P. Stone, "Autonomous Bidding Agents in the Trading Agent Competition," IEEE Internet Computing 5, no. 2, pp. 52-60, (2001).

[6]. Ana Cristina Bicharra Garcia, Anderson Lopes, Cristiana Bentes, Electronic Auction with autonomous intelligent agents: Finding opportunities by being there, Inteligencia Artificial, Revista Iberoamericana de Inteligencia Artificial. No.13 (2001), pp. 45-52. ISSN: 1137-3601. © AEPIA

[7]. Hong Liu, Real-Time Multi-Auctions and the Agent Support, Journal of Electronic Commerce Research, VOL. 1, NO. 4, 2000.

[8]. Tuomas Sandholm, Qianbo HuaiNomad," *Mobile Agent System for an Internet-Based Auction House*", IEEE Internet Computing, March/April 2000 (Vol. 4, No. 2). pp. 80-86.

[9]. Sproule & Archer, Knowledgeable Agents for Search and Choice Support in E-Commerce: A Decision Support Systems Approach, Journal of Electronic Commerce Research, VOL. 1, NO. 4, 2000

[10]. T. Bui and J. Lee, "An agent-based framework for building decision support systems," Decision Support Systems 25 no. 3, pp. 225-237, (Apr. 1999).

[11]. Hideyuki Mizuta Ken Steiglitz , Agent-Based Simulation of Dynamic Online Auctions – white paper.

[12]. Cong Chen Yiling, Chen Mark Cohen, Edward J. Glantz Rashaad E.T. Jones , Bidding Algorithm Comparison for Double Auctions, School of Information Sciences and Technology, The Pennsylvania State University, University Park, PA, IST 597B, Multi-Agent Systems SP03 Dr. Yen

[13]. A Roadmap of Agent Research and Development, Autonomous Agents and Multi-Agent Systems, 1, 275–306 (1998) c 1998 Kluwer Academic Publishers, Boston. Manufactured in The Netherlands.

[14]. Naoki Fukuta, Takayuki Ito, and Toramatsu Shintani, A Logic-based Framework for Mobile Intelligent Information Agents, Dept. Intelligence and Computer Science, Nagoya Institute of Technology, Nagoya, JAPAN..

[15]. Stuart Russell and Peter Norvig, Artificial Intelligence: A Modern, c 1995, Prentice-Hall, Inc.

[16]. Alper Caglayan and Colin Harrison, "Agent Source Book", John Wiley & Sons, Inc, United States of America, 1997.

[17]. Fabio Michael Wooldridge, Developing Multi-Agent Systems with JADE, Wiley Series in Agent Technology.

[18]. www.JADE.org.

[19]. Freeman Y. Huang, Heterogeneous Data Source Access in Web Applications, November 2000.

[20]. Prediction in Web site management using data mining – white paper.